# Online Route Choice Modeling for Mobility-as-a-Service Networks with Non-separable, Congestible Link Capacity Effects

Susan Jia Xu, Joseph Y. J. Chow

*Abstract*—With the prevalence of MaaS systems, route choice models need to consider characteristics unique to them. MaaS systems tend to involve service systems with fleets of vehicles; as a result, the available service capacity depends on the choices of other travelers in different parts of the system. We model this with a new concept of "congestible capacity"; that is, link capacities are a function of flow instead of link costs. This dependency is also non-separable; the capacity in one link can depend on flows from multiple links. An offline-online estimation method is introduced to capture the structural effects that flows have on capacities and the resulting impacts on route choice utilities. The method is first applied to obtain unique congestible capacity shadow prices in a multimodal network to verify the capability to capture congestion effects on capacities. The capacities are shown to vary and impact the utility of a route. The method is validated using real system data from Citi Bike in New York City. The results show that the model can fit to the data quite well and performs better than a baseline modeling approach that ignores congestible capacity effects. By relating the route choice to congestible capacities using a random utility model, modelers can monitor and quantify the impacts to traveler consumer surplus in real time. Applications of the model and online method include monitoring capacity effects on consumer surplus, using the model to direct incentives programs for rebalancing and other revenue management strategies, and to guide resource allocation to mitigate consumer surplus impacts due to disruptions from incidents.

## I. INTRODUCTION

TRAVELER information has exploded over the past decade with the development and use of Intelligent Communication Technologies (ICTs) to detect and analyze traffic conditions. The up-to-the-minute information is provided in many places with travel websites, real-time roadside infrastructure, "next-bus" displays, etc., which change when, where, and how we travel *[1]*, and has only expanded in recent years with mobile devices and associated mobility services. For example, not only can travelers choose to drive, take subway or bus, bike, walk, or take taxi; they also have a host of other mobility options accessed via mobile device apps: station-based and dockless shared bikes, shared taxi options like Via, Uber, or Lyft, and car sharing options like ReachNow or Car2Go. Driven by data and technology, similar types of transport services can be found in many major cities around the world. The ecosystem for urban mobility is swiftly changing from one of car-dependency to a more multimodal *[2]*, Mobility-as-a-Service (MaaS) *[3 – 5]* setting.

Operation of MaaS systems face challenges associated with traveler information. In such systems, travelers inherently interact with the MaaS by accessing some type of vehicular or micromobility service in real time to make trips *[6]*. Example providers like MaaS Global, Masabi, Moovit, and their integration of different operators onto one platform come to mind. The online nature of these services requires operators to make dynamic fleet decisions like rebalancing idle vehicles *[7, 8]*, updating prices and vehicle routes *[9]*, updating vehicle schedules *[10]*, among others. These operations depend on accurately measuring their impacts on traveler demand at the route choice level. For example, having a certain number of idle shared bikes docked at a station should impact travelers' choices of where they pick up or drop-off their bikes. In other words, MaaS platforms require effective route choice models to inform operators and provide decision support for their dynamic operations.

However, route choice models for highly dynamic multimodal networks face unique challenges that need to be overcome, especially with emerging technologies *[11]*. Travelers use mobile ticketing and reservations to pick up or board a vehicle trip at one multimodal facility to get to another *[6]*. As a result, the capacity of vehicles (spaces) at stations for pick up or boarding (drop off or alighting) are dynamic and depend on inbound and outbound flows of other travelers *[7]*.

Unlike conventional traffic networks in which the link travel cost may exhibit congestion effects with link performance functions (see Bell *[56]*), MaaS systems have link capacities that exhibit congestion effects. The link capacity of this system can represent, for example, availability of vehicles or passenger space during time interval $t$. In the case of bikeshare or carshare system, the number of available bikes is the pickup link (virtual link at a station connecting a walk link to a bike link) capacity

Submitted on Jan. 16, 2021, for review. This study was conducted with support from the NSF grants CMMI-1634973 and CMMI-1652735.

Susan Jia Xu was with the New York University, New York, NY 11201 USA (e-mail: jx731@nyu.edu). She is now with the San Diego Association of Government (SANDAG), San Diego, CA 92101 USA.

Joseph Y.J. Chow is with the Department of Civil & Urban Engineering, New York University, New York, NY 11201 USA (e-mail: joseph.chow@nyu.edu).

and the number of empty bike docks is the drop-off link capacity. The link capacity may be the number of passengers that can be transported by a vehicle in microtransit, or number of passengers per hour for fixed route transit with a line capacity. In this study we call this *congestible capacity* (and to the best of our knowledge has not been studied yet). The capacity effect now also depends on that combined interaction of operator policies and travel choices. If the balance of travelers bringing vehicles to a location versus taking vehicles away from a location changes between time periods, it should impact the capacity and its effect on travelers' route choices. Similarly, if an operator were to change their rebalancing algorithm parameters for a different time of day, it should be reflected by a different system efficiency matrix estimated offline.

Because these congestible capacities at each link are influenced by multiple inbound and outbound flows from other links, it results in non-separable (see *[12]*) link capacities that depend on multiple links flows. Lastly, first-in-first-out queueing characteristics also exist in such systems; for example, even if an initially empty facility had 10 vehicles dropped off in an hour, it does not necessarily mean that there is sufficient capacity available for 10 individuals arriving within that same hour as it depends on when they arrive. As a result, the effects of capacity for a given time interval are not straightforward to assume, and therefore need to be treated as latent, unobservable variables during that same time interval *[13]*.

Despite its long history *[14 – 19]*, route choice models have not considered latent congestible capacity effects to model route choice in dynamic multimodal networks with online information. Prior work on multimodal network route choice focused on choice set generation considering overlap (e.g. *[20 – 23]*). Another related research area is route choice under real time information (e.g. *[24 – 28]*), but that work focused on user perceptions and equilibration/adaptation/learning with regards to information provision. Other related research on traffic assignment problem have dealt with capacity constraint. However, they have not delved into modeling the dynamics of system congestion effects on capacitated route choice nor on congestible capacities.

We propose an online route choice model, updated each time interval from prior time interval data, to provide a forecast of route choices for the subsequent interval. The estimation and updating of such a model allows a MaaS operator to monitor the effects of their dynamic decisions on users' route choice behavior. This is done by introducing a latent congestible capacity variable to capture the dynamic capacity effects in multimodal systems and using observations from prior interval(s) to forecast the latent variable and its effect on dynamically capacitated route choice.

The reminder of the paper is organized as follows. Section II presents the online system setting and a literature review to addressing the route choice model. Section III presents the proposed online route choice model. Numerical experiments are conducted using synthetic data in Section IV to verify the model. Section VI presents a case study based on historical trip data from the New York City (NYC) bike-sharing system, Citi Bike. Finally, the conclusions are drawn, and future research is discussed.

## II. PROBLEM DESCRIPTION

Modeling route choice behavior is essential to forecast travelers' behavior under hypothetical scenarios and to understand travelers' reaction and adaptation to sources of information *[29]*. Present research directions show growing interest in understanding travelers' behavior under multimodal networks (e.g. *[30 – 33]*). To model route choices in a multimodal network, one needs an extensive representation of valuations and preferences that individuals have regarding attributes of route components *[30]*. A well-known multimodal transport network scenario is Park-and-Ride, which provides parking facilities at the edge of city centers to encourage parking and transfers to public transit *[34 – 38]*. New shared mobility systems, such as bikeshare, are involved as first/last mile transport modes *[39]*.

We consider an online system in which a MaaS operator receives link flow information $x_{at}$ for each link $a \in A$ in time interval $t$ with the network denoted by a directed graph $G(N, A)$. The set of links may be further divided into different modes (e.g. walking, transfers in station, or in-vehicle) as $\boldsymbol{A} = \{A_0, A_1, \ldots, A_M\}$, where the index 0 denotes walking. The system may be scheduled or on-demand (in which case it is a Mobility-on-Demand system). Each link has a fixed, generalized cost $c_a$ and a capacity $u_{at}(\boldsymbol{x_t}), a \in A$, which varies with time interval $t$ because of a vector of non-separable link flows, $\boldsymbol{x_t} = \{x_{1t}, \ldots, x_{|A|t}\}$. The $\boldsymbol{u_t}$ can only be measured after the interval $t$.

The common characteristic assumed for this system is that the capacity varies sufficiently dynamically within an interval that the precise value perceived by travelers during the same time interval varies by traveler and is not perfectly observable to the system. An overview of the system is shown in **Fig. 1**.

We focus on the route choice model $\Omega_t$. A traveler in period $t$ makes a choice of route $k \in K_{rst}$ on the directed graph $G(N, A)$ to get from origin $r$ to destination $s$, $(r, s) \in W$, with probability $P_{rst}(k)$. The choice depends on availability of the congestible capacity, which in turn depends on the flows throughout the network (including the traveler whose route choice is being modeled). A methodology is needed to estimate $\boldsymbol{u_t}(\boldsymbol{x_t})$ and determine $P_{rst}(k)$ in such a system.

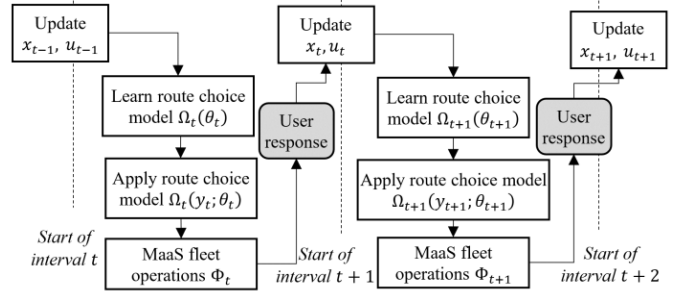

Fig 1. Illustration of functions operating within an online MaaS.

Capacity effects on route costs can be modeled using shadow prices $\boldsymbol{w_t}(\boldsymbol{x_t})$ *[40, 41]* which capture the opportunity cost of the system having the capacity bound at a given amount and resulting in diverting travelers to next-best alternatives. These values are unobservable. Since these shadow prices depend on the choices of other travelers $(x)$, these can be modeled similarly to the "field effect" observed in social influence-based discrete choice models *[42 – 44]*. For example, Walker et al. *[45]* presented a utility function with social influence as shown in Eq. (1), where $V(q_{in}, Z_n; \beta)$ is the systematic utility of individual-specific attributes $q_{in}$ and alternative-specific attributes $Z_n$, $F_{in}$ is an endogenous proportion of people in the decision-maker's peer group choosing alternative $i$, and $\varepsilon_{in}$ is a random disturbance. The parameter $\gamma$ reflects the influence that the peer group has on the individual's choice with positive values demonstrating a bandwagon-type of effect while negative values demonstrating congestion effects.

$$U_{in} = V(q_{in}, Z_n; \beta) + \gamma F_{in} + \varepsilon_{in} \qquad (1)$$

Endogenous effects have been studied extensively in recent years *[44 – 46]* but they mostly pertain to social network effects on long term decisions like technology adoption or mode choice. Furthermore, the methodology revolves around using a linear "field" effect. In the context of capacitated route choice, a linear effect ignores the nonlinear dynamics resulting from multiple link flows interacting on a link's capacity. These studies, however, offer insights on how to handle the endogeneity. One way is by estimating the field effect in a two-step approach.

In addition to unobservable link capacity effects to exhibit endogeneity due to congestion effects, the effects can depend on multiple other link flows, i.e. non-separability. In the literature, non-separability is typically applied to link costs (e.g. *[47, 48]*), not link capacity effects. The link capacity shadow price can be estimated dependent on link flows under the setting of an online system.

To summarize, the problem involves congestible capacity effects that are typically unobserved but dependent nonetheless on decisions of other travelers. Based on the literature, we consider using shadow prices to capture the unobservable capacity effect and handle its endogeneity by using a two-step estimation approach to estimate capacity from link flow patterns obtained from the prior time interval and follow that with shadow price estimation.

III. PROPOSED METHODOLOGY

With respect to the route choice modeling, the two main steps are (1) generating a realistic set of choices (routes); and (2) modeling route choice given such a set of choices. Since the focus of this study is on the estimation of the route choice with congestible link capacities, we assume that the route choice set is provided. For studies on choice set generation readers are referred to Prato *[29]*.

*A. Model formulation*

A random utility model (RUM) is formulated as follows. For a given choice set $\boldsymbol{K_{rst}}$, the utility of a route $k \in \boldsymbol{K_{rst}}$ is composed of links $a \in A_{km}$, where $A_{km} \subseteq \boldsymbol{A_m}$ is the set of links forming the portion of route $k$ in mode $m$, with generalized costs $c_a$ for traveler in a period $t$ (unit of observation) as defined in Eq. (2). The attribute $w_{at}$ is the shadow price corresponding to the link capacity $u_{at}$ and is a function of the set of link flows $x_t$. The parameter $\theta_m$ is used to scale the degree of dispersion in perception of the travel cost differences for each modal link, where a higher value corresponds to less indifference between two routes with different travel costs.

$$U_{k,rst} = -\sum_{m=0}^{M} \theta_m \sum_{a \in A_{km}} (c_a + w_{at}(x_t)) + \varepsilon_{k,rst} \qquad (2)$$

where $V_{k,rst} = -\sum_{m=0}^{M} \theta_m \sum_{a \in A_{km}} (c_a + w_{at}(x_t))$ is the representative utility and $U_{k,rst}$ is the utility of route $k$ (the altnerative $i$ in Eq. (1)) for users of OD $(r,s)$ in period $t$ (as individual $n$). When $\varepsilon_{k,rst}$ is Gumbel distributed, the conditional logit form is shown in Eq. (3). Note that more complex expressions can be considered, but we stick to a basic multinomial logit (MNL) formulation for the sake of better interpreting the relationship with the congestible capacity.

$$\Pr(k, rst) = \frac{\exp(V_{k,rst})}{\sum_{k' \in K_{rsn}} \exp(V_{k,rst})} \qquad (3)$$

The shadow prices $w_{at}(x_t)$ are unobservable, but through network flow complementary slackness conditions (see *[41]*) we know that the properties in Eq. (4) must hold.

$$if\ x_{at} = u_{at}, w_{at} \geq 0 \qquad (4a)$$

$$if\ x_{at} < u_{at}, w_{at} = 0 \qquad (4b)$$

We can further relate the capacities $u_{at}$ to the link flows. Depending on arrivals of travelers to a link, the capacity will vary. This random capacity is estimated using data from the prior time interval to determine the effective coefficients associated with all inbound and outbound link flows such that the resulting $u_{at}$ and $w_{at}$ fit best. We define a linear system efficiency matrix similar to input-output models (see Leontief *[55]*) for statistically describing $\boldsymbol{u_t} = f(\boldsymbol{x_t}, \boldsymbol{u_{t-1}})$, where the parameters relate to the technical efficiency observed in converting flows into impacts on capacity for a given time interval. Different system operating policies (e.g. rebalancing, routing) will impact the parameters. For a given link $a \in \boldsymbol{A_m}$ in a mode $m$, there is a set of inbound link flows $I_T(a)$ and outbound link flows $O_T(a)$ at the tail node. There is also a set of inbound link flows $I_H(a)$ and outbound link flows $O_H(a)$ at the head node. The capacity is forecasted with a set of simultaneous equations shown in Eq. (5).

$$\hat{u}_{at} = u_{a,t-1} + \beta_m^{I_T} \sum_{a' \in I_T(a)} x_{a't}$$
$$- \beta_m^{O_T} \sum_{a' \in O_T(a)} x_{a't}$$
$$- \beta_m^{I_H} \sum_{a' \in I_H(a)} x_{a't} \qquad (5a)$$
$$+ \beta_m^{O_H} \sum_{a' \in O_H(a)} x_{a't} + \gamma_{at},$$
$$\forall a \in A_m, 0 \leq m \leq M$$

Subject to

$$x_{at} \leq u_{at}, \qquad \forall a \in A_m, 0 \leq m \leq M \quad (5b)$$

where $\gamma_{at}$ is a random disturbance term across each observation $t$, assuming $\gamma_{at} \sim N(0, \sigma_a^2)$. There should be $4 + t$ parameters and $|A_m|t$ equations for one mode $m$. The $4 + t$ parameters are associated with the inbound and outbound link flows of head and tail nodes with a given mode $m$, and one parameter $\sigma_a$ for the link-specific random disturbance term.

The $u_{a,t-1}$ and $u_{at}$ are the capacities observed at the end of the preceding period and the current period, respectively. The signs preceding the parameters reflect the general effect of having vehicle capacity versus space capacity. The values of the parameters $\boldsymbol{\beta}$ should be between -1 and 1, where a value of 1 implies perfect efficiency in transferring the vehicle flows into or out of capacity for the link, i.e. all inbound capacity arrives first before all outbound demand. In dynamic systems the randomness of the arrivals impacts the effective value of $\boldsymbol{\beta}$, similar to how arrival patterns at traffic signals impact the effective capacity of the approach.

Eqs. (2) – (5) are related as follows. The values of $\beta_m^{IT}, \beta_m^{OT}, \beta_m^{IH}, \beta_m^{OH}, \sigma_a$ in Eq. (5) should be estimated offline to capture typical structure of arrival patterns. Eq. (5) and the $x_{at}$ values determine the value of each $u_{at}$. The values of $w_{at}$ are estimated using maximum likelihood constrained to the values of $u_{at}$ in Eq. (4). The representative utilities $V_{k,rst}$ can then be specified with the $w_{at}$ to determine the $P_{rst}(k)$. The unimodal network in **Fig. 2** is used to illustrate the equations. This network has two nodes having four paths (closed system) as links. Eqs. (5) are simplified into Eqs. (6).

$$\begin{aligned} u_{at} &= u_{a,t-1} + \beta_1^{IT} x_{ct} - \beta_1^{OT} x_{at} - \beta_1^{IH} x_{at} \\ &\quad + \beta_1^{OH} x_{ct} + \gamma_{at} \\ &= u_{a,t-1} + \beta_1 x_{ct} - \beta_2 x_{at} + \gamma_{at} \end{aligned} \quad (6a)$$

$$\begin{aligned} u_{bt} &= u_{b,t-1} + \beta_2^{IT} x_{dt} - \beta_2^{OT} x_{bt} - \beta_2^{IH} x_{bt} \\ &\quad + \beta_2^{OH} x_{dt} + \gamma_{bt} \\ &= u_{b,t-1} + \beta_3 x_{dt} - \beta_4 x_{bt} + \gamma_{bt} \end{aligned} \quad (6b)$$

$$\begin{aligned} u_{ct} &= u_{c,t-1} + \beta_3^{IT} x_{at} - \beta_3^{OT} x_{ct} - \beta_3^{IH} x_{ct} \\ &\quad + \beta_3^{OH} x_{at} + \gamma_{ct} \\ &= u_{c,t-1} + \beta_5 x_{at} - \beta_6 x_{ct} + \gamma_{ct} \end{aligned} \quad (6c)$$

$$\begin{aligned} u_{dt} &= u_{d,t-1} + \beta_4^{IT} x_{bt} - \beta_4^{OT} x_{dt} - \beta_4^{IH} x_{dt} \\ &\quad + \beta_4^{OH} x_{bt} + \gamma_{dt} \\ &= u_{b,t-1} + \beta_7 x_{bt} - \beta_8 x_{dt} + \gamma_{dt} \end{aligned} \quad (6d)$$

Subject to
$$x_{at} \leq u_{at}, x_{bt} \leq u_{bt}, x_{ct} \leq u_{ct}, x_{dt} \leq u_{dt} \quad (6e)$$

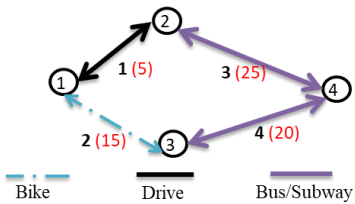

Fig 2. The simple multimodal network used for illustrating the methodology

### B. Offline-online estimation

The parameters $\theta_m$ capture the perceived effects of modal travel costs such as for walk time, in-vehicle time, fare price, or binding capacity as shown in Eq. (2) (see Bell [56]). As $\theta_m \to 0$, travelers become indifferent between routes, whereas $\theta_m \to \infty$ suggest a deterministic perception of travel time. These travel behavioral parameters are estimated offline from historical route choice data during uncongested period set $S_1$ where $w_{at} = 0, n \in S_1$. The basis of the offline-online estimation is the following logic: since $\theta_m$ are time-invariant, the estimated values can then be used during the congested setting to estimate the $w_t$ in an online setting without bias. For a RUM, the value of the parameters can be estimated by maximizing the log-likelihood function in Eq. (7), where $y_{kt} = 1$ if an observed trip between $(r, s)$ chooses route $k \in K_{rs}$. Since this is offline, the time period $t$ can be left out of the observation indexing. For logit models with Gumbel disturbances, Eq. (7) is concave [54].

$$\hat{\theta} = \arg\max_\theta \left\{ \sum_{t \in S_1} \sum_{(r,s) \in W} \sum_{k \in K_{rst}} y_{kt} \ln P_{rst}(k; \theta) \right\} \quad (7)$$

The parameters $\beta_m^{IT}, \beta_m^{OT}, \beta_m^{IH}, \beta_m^{OH}, \sigma_a$ capture the efficiency of the network structure, similar to how traffic models have parameters capturing the platooning characteristics of arrival patterns and input-output models in macroeconomics describing technologies in converting one commodity to another. These parameters are also estimated offline from historical data from a period set $S_2$ on Eq. (5a), e.g. inflows and outflows and the resulting average capacities, once for each mode $m$. In the offline estimation, the capacities $\boldsymbol{u_t}$ are observed as the time-average capacities during the time interval $t$. Each parameter and disturbance are assumed independent of the other equations. As such, we estimate each equation as an independent linear regression model with multiple post-interval observations of capacity and inflows/outflows using ordinary least squares.

The remaining $w_{at}$ need to be updated in the online system each interval $t$. We estimate the values of $w_{at}$ for the current time interval using constrained maximize likelihood as shown in Eq. (8) based on $\hat{\boldsymbol{u}}_t$ forecast from the prior time interval. The constraints in Eq. (8b) are set from $\hat{\boldsymbol{u}}_t$ and using $\boldsymbol{x}_{t-1}$ as an approximation $\hat{\boldsymbol{x}}_t$. By design, the constrained optimization ensures that a unique set of shadow prices are found: even in the case where there may be multiple $w_{at}$'s that fit the observed flow, Eq. (8a) is a continuous, concave function of $w_{at}$ and will lead to an optimal value. Estimation of $w_{at}$ is done by first setting $\hat{w}_{at} = 0$ if $\hat{x}_{at} = \hat{u}_{at}$ and then estimating the remainder via Eq. (8a) while requiring $\hat{w}_{at} \geq 0$.

$$\hat{w}_t = \arg\max_{w_t} \left\{ \sum_n \sum_{(r,s) \in W} \sum_{k \in K_{rst}} y_{kn} \ln P_{rst}(k; w_t | \hat{u}_t, \hat{\theta}, \hat{x}_t) \right\} \quad (8a)$$

Subject to
Eqs. (4) with $\hat{\boldsymbol{u}}_t, \hat{\boldsymbol{x}}_t$ $\quad (8b)$
$$\hat{w}_t \geq 0 \quad (8c)$$

The complete offline estimation and online system learning process is summarized in **Algorithm 1**.

**Algorithm 1. Offline-online estimation and system learning**

OFFLINE ESTIMATION
1. Estimate $\hat{\theta}_m$ using Eq. (7) on sample route choice data under uncongested conditions where $\boldsymbol{w_t} \equiv 0, t \in S_1$.

2. Estimate $\{\hat{\beta}_m^{I_T}, \hat{\beta}_m^{O_T}, \hat{\beta}_m^{I_H}, \hat{\beta}_m^{O_H}, \hat{\sigma}_a\}$ using Eq. (5a) on historical operational data with post-interval values of $u_t$ for each interval $t \in S_2$.

ONLINE LEARNING: at the start of each interval $t$,

3. Observe the values of $x_{t-1}, u_{t-1}$.
4. Update $\hat{u}_t$ using Eq. (5) and assume $\hat{x}_t = x_{t-1}$.
5. Update the shadow prices $\hat{w}_t$ using Eq. (8).

Once values of $\hat{w}_t$ are estimated, we can use the model $P_{rsn}(k|\hat{\beta}_m^{I_T}, \hat{\beta}_m^{O_T}, \hat{\beta}_m^{I_H}, \hat{\beta}_m^{O_H}, \hat{\sigma}_a, \hat{w}_t)$ to determine route choices for each OD pair $(r,s)$ in time interval $t$. The model is validated at two levels. The first is at the offline level to ensure that $\hat{\theta}_m$ and $\{\hat{\beta}_m^{I_T}, \hat{\beta}_m^{O_T}, \hat{\beta}_m^{I_H}, \hat{\beta}_m^{O_H}, \hat{\sigma}_a\}$ fit the data from out-of-samples from $S_1$ and $S_2$. The second is at the online level. Since capacities and link flows are observed after the end of each period $t$, if we have sampled observations of route choices each period we can then validate the performance of the forecast model.

*C. Discussion of model properties*

This model, applied over time, provides a monitor of the traveler behavior and can be used to measure the impacts of any system changes on changes in behavior. Example uses of this monitoring include the following operational use cases:
- Identifying thresholds in link capacity shadow prices where route choice elasticities are of interest;
- Identify thresholds in link volumes in which case the congestion impacts on link capacities are critical;
- Online revenue management strategies like incentivizing travelers to switch routes or directing service staff or vehicles to mitigate critical capacities;
- Identify critical nodes in the network and over multiple time periods in which the route choices most impact the system performance throughout the network.

Each of the model estimation steps have their corresponding goodness-of-fit measures as discussed in the prior section. Evaluation of the online system overall is done using post-interval comparison of predicted route flows and realized route flows over multiple time periods. Since the application is an online system, no flow equilibration (see *[49, 50]*) needs to be assumed.

Eq. (5a) assumes normally distributed disturbances. In future studies we will explore extreme value distributions like a Weibull distribution which may better reflect the maximum value distribution of capacity.

The methodology for estimating $\hat{w}_{at}$ assumes a myopic approach to the online learning, using $\hat{x}_t = x_{t-1}$. This can be problematic especially under larger time intervals. One way that might improve this estimate is to model the longitudinal behavior of the $x_n$ using a time series model $\hat{x}_t = f(x_{t-1}, x_{t-2}, ...)$ (see *[51]*) and apply that model to forecast the flows in the current time interval. That will also be studied in the future.

IV. VERIFICATION EXPERIMENTS

The proposed methodology is first verified in this section to show that it works as intended. Two numerical experiments are conducted using multimodal networks under an offline and online system, respectively. The experiments have two primary objectives.

The first objective is to demonstrate the flexibility of the proposed method to adapt to links where capacities are binding in some periods. This is accomplished by applying the estimated models to compute route choice probabilities in observed time periods.

The second objective is to show the ability of the proposed method in detecting changes in $w_{at}$ due to demand changes.

The validation of the methodology is conducted in Section IV.

*A. Multimodal network in offline system*

The first numerical experiment is conducted on a MaaS network with congestible capacity effects on multiple links in an offline system, where each observation period assumes $x_t$ and $u_{t-1}$ are known. Since there's no separate offline data set without congestion, we simply assume the simulated ground truth degree of dispersion $\theta$ in this example is known: $\theta = 0.0905$. Consider a network as shown in **Fig. 2** with four links, where each link corresponds to one type of transport mode. The generalized travel time for each link is shown in red parentheses.

The travelers' route choices are generated randomly for 100 independent observations in 100 independent time intervals. The sampled data set is available on our Github site *[52]*. There are two paths in the choice set represented by the following link sequences: (1,3), (2,4), where their generalized travel costs are 30 and 35, respectively. In this test, there are congestible capacity effects observed on facility node 1, 2, and 3. Travelers who choose path 1 (1,3) are constrained by the number of available parking spaces at facility node 2. Travelers who choose path 2 (2,4) are restricted to the number of shared bikes for pick up at node 1 and the number of open docks for drop off at node 3.

Firstly, it is important to illustrate the capability of the proposed method to capture the heterogeneity of different travelers' effects. Eqs. (9) describe the general effect of flows into or out of each link on the capacities. The sign "-" stands for the traveling direction from node 1 to node 4, and the "+" is the opposite direction. There are two capacity functions for the bike link, since each bike station has bikes for pickup and docks for dropping off. The proposed method sets the capacity as a function of the observed flow to and from that facility. Hence, the congestible capacity functions are formulated as Eq. (9).

$$m = 1^- \quad u_{1^-,t} = u_{1^-,t-1} - \beta_1 x_{1^-,t} + \beta_2 x_{1^+,t} + \gamma_{1^-,t} \quad (9a)$$
$(drive)$

$$m = 2^- \quad u_{2p^-,t} = u_{2p^-,t-1} + \beta_3 x_{2^+,t} - \beta_4 x_{2^-,t} + \gamma_{2p^-,t} \quad (9b)$$
$(bike)\ pick\ up$

$$m = 2^- \quad u_{2d^-,t} = u_{2d^-,t-1} - \beta_5 x_{2^-,t} + \beta_6 x_{2^+,t} + \gamma_{2d^-,t} \quad (9c)$$
$(bike)\ drop\ off$

$$m = 2^+ \quad u_{2p^+,t} = u_{2p^+,t-1} + \beta_7 x_{2^-,t} - \beta_8 x_{2^+,t} + \gamma_{2p^+,t} \quad (9d)$$
$(bike)\ pick\ up$

$$m = 2^+ \quad u_{2d^+,t} = u_{2d^+,t-1} - \beta_9 x_{2^+,t} + \beta_{10} x_{2^-,t} + \gamma_{2d^+,t} \quad (9e)$$
$(bike)\ drop\ off$

The parameters are estimated via regression of Eq. (5) across the 100 observations. The parameters are estimated in **Table 1**. In general, the inbound flows increase the capacity (e.g. vehicles return) and the outbound flows decrease the capacity. The signs preceding the parameters reflect the general effect of having vehicle/bicycle capacity versus space capacity, which are expected. The values of the parameters $\beta$ are expected to be between 0 and 1. The magnitude of the parameters $\boldsymbol{\beta}$ of the drive link are higher than ones of bike link, which suggests the drive link has higher efficiency in transferring the vehicle flows into or out of capacity. Once the $\boldsymbol{\beta}$'s are estimated, the link capacities $u_t$ for each time interval are computed as $u_t = u_{t-1} + \boldsymbol{\beta}^T \boldsymbol{x}_t$. The predicted capacity on the bike link $(2^-)$ across the 100 time intervals is shown in **Fig. 3**.

Table 1. Parameters estimation results

|  | $x_1+$ | $x_1-$ | $x_2+$ | $x_2-$ |
|---|---|---|---|---|
| $m = 1^-$ (drive) | $\beta_2$:0.5526 | $\beta_1$:0.6636 |  |  |
| $m = 2^-$ (bike) pick up |  |  | $\beta_3$:0.3959 | $\beta_4$:0.2964 |
| $m = 2^-$ (bike) drop off |  |  | $\beta_6$:0.5020 | $\beta_5$:0.3759 |
| $m = 2^+$ (bike) pick up |  |  | $\beta_8$:0.2710 | $\beta_7$:0.2029 |
| $m = 2^+$ (bike) drop off |  |  | $\beta_9$:0.3570 | $\beta_{10}$:0.2673 |

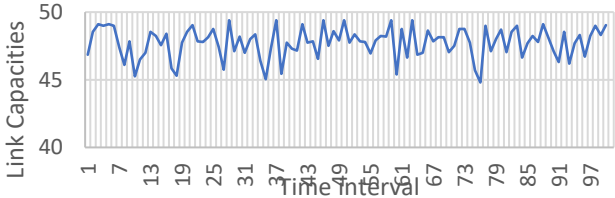

Fig. 3. The predicted link capacity of bike pick-up link (u2p-) across the 100 time intervals.

### B. Multimodal network in online system

The example from Section IV.A is continued now in an online context. Based on this value and the results from IV.A, we updated the shadow prices $w_{at}$ using Eq. (8). The travel time in the utility function is updated to the sum of constant link cost and the value of the shadow prices. The route choice probability can then be computed, as illustrated for path 2 in **Fig. 4**. We verified that the proposed model is flexible enough to adapt to links where capacities might by binding in some periods and not in others in a multimodal network.

Since the demand plays an important role in the proposed model, we need to determine how the effect of congestible capacity varies with the demand. The comparison among different scenarios is shown in **Fig. 5**. For the 101st observation, there is a higher demand 105 from node 1 to node 4, and observed path flows are {58,47}. The shadow price of path 1 is estimated to 2.68 and its probability is calculated as 0.55. For the next observation, the demand from node 1 to node 4 is decreased to 95. We run the proposed model, and the new shadow price of path 1 is estimated to 0.03. The probability of path 1 is increased to 0.61. Because path 1 is the shortest when no capacity is considered, the path probability would always decrease from the uncapacitated state since the capacity would increase the effective cost relative to path 2. We can quantify how a higher demand results in higher $w$ and lower probability to choose path 1. Furthermore, this quantified congestion effect is related to other flows in the network. For example, path 2 shows fluctuations in path probability during observations 1 to 40, so during that time the decision-maker can look to strategies that impact bike drop-off, i.e. Eq. (9c), and note that the capacity is 34% more sensitive to inbound flow $x_2+$ (0.5020) than to outbound flow $x_2-$ (0.3759).

### V. MODEL VALIDATION: CITI BIKE IN NEW YORK CITY

The proposed model is tested using trip historical data from *Citi Bike* – the unique bikesharing system in New York City (*citibikenyc.com*). Since the purpose of this experiment is to validate the effectiveness of the model using real data, we can simply rely on a bikeshare subnetwork in which travelers use two modes to get to their destinations: walk and bike. The setting includes multiple modes (walk from origin to bike station pickup, bike to station drop-off, walk to destination) and interacting flows (the effects of external flows are accounted for in the estimation of the $\boldsymbol{\beta}$'s in the offline estimation).

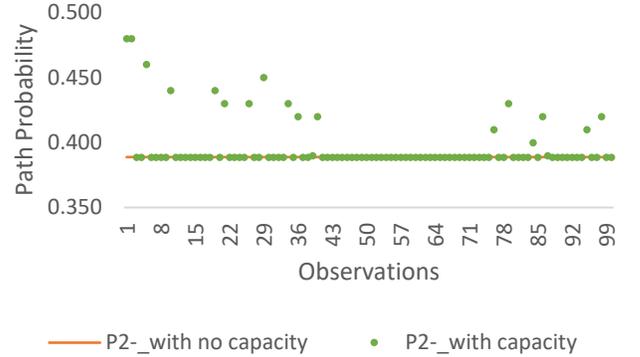

Fig. 4. Comparison of path 2 probabilities in the multimodal network.

### A. Data and experiment design

A subnetwork is extracted from the whole *Citi Bike* service system as shown in **Fig. 6** overlaid upon a Google Maps image. The zones in the study area are categorized by *Census 2010* (see Open Data in Department of City Planning in NYC). The centroid of the zone is created to represent origin or destination for travelers making trips from or to the zone. This is consistent with conventional transportation network modeling practice, where the zone sizes and centroids are selected to properly represent population clusters. If the walk trips were removed and only station-to-station considered, farther apart station pairs that are situated closer to centroids would confuse the model. Aggregated demand for each Census Tract (CT) by time interval were generated from *Citi Bike* historical data. The network is designed to have 17 zones and 41 bikesharing stations. While predicting traveler's choice, they are assumed to pick up and drop off bikes to the nearest station. The general travel cost is the sum of walking time from the zone centroid to the pickup station, cycle time, and the walking time from the drop-off station to the zone centroid.

Prior to process bikesharing trip data, we checked the weather data for the month of July in 2018. Dates with clear and good weather are preferred, because of the control of environmental variables effect on user's choice. Moreover, the aggregate ridership for each weekday (e.g. Monday to Friday) in July 2018 is checked. For five consecutive weekdays, the daily ridership should not be too different. Hence, five weekdays of *Citi Bike* trip historical data from July 9th, 2018 to July 13th, 2018 are used as a test data set, and one day of trip historical data on July 18th, 2017 is used as the training data set. The observed time interval is set to 30 minutes, as *Citi Bike* membership include unlimited 30-min rides. The travel cost is computed as distance divided by speed. The following steps are taken to prepare data for the proposed route choice model learning method, and the sample of a data frame for one OD within a time interval is shown in **Table 2**.

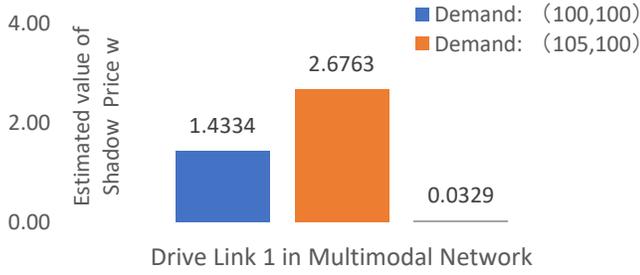

Fig. 5. Comparison of congestible capacity effect on drive link due to demand change.

Table 2. A sample of required data frame for the one OD within a time interval

| Start CT | End CT | start.station id | end.station id | choice | cost | infreq | outfreq | out demand | in demand |
|---|---|---|---|---|---|---|---|---|---|
| 13100 | 10100 | 447 | 379 | 0 | 12.66 | 1 | 1 | 7 | 2 |
| 13100 | 10100 | 447 | 3255 | 0 | 15.08 | 3 | 1 | 7 | 2 |
| 13100 | 10100 | 447 | 492 | 1 | 12.13 | 3 | 1 | 7 | 2 |
| 13100 | 10100 | 447 | 490 | 0 | 14.5 | 0 | 1 | 7 | 2 |
| 13100 | 10100 | 469 | 490 | 0 | 13.06 | 0 | 0 | 7 | 2 |
| 13100 | 10100 | 469 | 379 | 0 | 11.04 | 1 | 0 | 7 | 2 |
| 13100 | 10100 | 469 | 3255 | 0 | 13.63 | 3 | 0 | 7 | 2 |
| 13100 | 10100 | 469 | 492 | 0 | 10.5 | 3 | 0 | 7 | 2 |
| 13100 | 10100 | 500 | 492 | 0 | 9.33 | 3 | 1 | 7 | 2 |
| 13100 | 10100 | 500 | 490 | 0 | 11.9 | 0 | 1 | 7 | 2 |
| 13100 | 10100 | 500 | 379 | 0 | 9.87 | 1 | 1 | 7 | 2 |
| 13100 | 10100 | 500 | 3255 | 0 | 12.48 | 3 | 1 | 7 | 2 |

The magnitudes of the shadow prices give a relative measure of the insufficient capacity in the link with respect to other links. Finally, three different route choice models are applied to make qualitative comparisons:

(i)  MNL with generalized travel cost;
(ii) MNL with updated travel cost, where shadow prices are estimated under constant capacity;
(iii) MNL with updated travel cost, where shadow prices are estimated under constraints determined by congestible capacity function.

Note that the subnetwork is subject to flows to/from external stations, which are not modeled using gateway stations since there would be other sources of noise like distances between external centroids and those stations. Since the objective is to compare across models in the same environment, this should remain a fair comparison. Furthermore, while the system efficiency matrix does not directly include the external flows, any system increases or decreases would be accounted for in the coefficients at stations near the edges or be captured by the error terms. For the same reason, a single value of theta is used for both the bike and walk modes.

### B. Citi Bike system case study results

The test data set is used to estimate parameters in the congestible capacity function. Since we set 30 minutes as one observation period, there are 240 time intervals in total for the five consecutive weekdays data set. The one-day training data set includes 48 time intervals. Each station has two congestible capacity functions, one for pickup and another for drop-off. Given specifications and observations of different time intervals with network flows and initial capacities at the start of each observed period, the parameters of congestible capacity function are estimated. **Fig. 7** illustrates the performance of the proposed method to estimate congestible capacity to station level over time. It shows in-sample and out-of-sample trajectories of the capacities for the station #519, which has the highest in and out frequencies in the study network (location shown in **Fig. 6** is at Grand Central Station). Because of the assumption of lag in the update of the flow (step 4 of Algorithm 1), naturally parts of the estimated versus observed will look like there is a lag. Major differences that occur would indicate a strong shift in the flow patterns. The full results of all other stations are shown in Xu *[57]* and posted as Appendices on the Github site *[52]*, which includes the p-values for the estimated $\beta$'s in Appendix D.

In **Fig. 8**, the upper plot shows a comparison between in-sample capacity predictions and observations over five consecutive weekdays for station #519, and the lower one

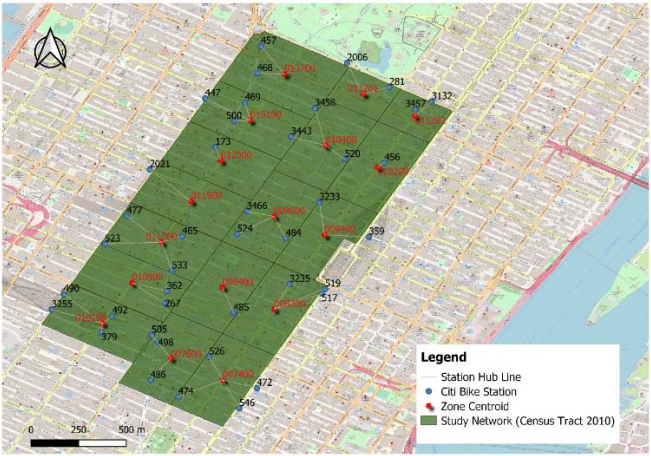

Fig. 6. Study network in the case study.

1. To extract trip historical data from *Citi Bike System Data* on the following days: 07/09/2018, 07/10/2018, 07/11/2018, 07/12/2018, 07/13/2018, and 07/18/2018.
2. To import the map of study network (e.g. 17 Census Tract (CT) zones)
3. To filter *Citi Bike* trip historical data by study area (e.g. *Citi Bike* stations that are included in the study network, see **Fig. 6**)
4. To aggregate *Citi Bike* trip data by time interval and CT zone
    a. Trip data by time intervals
    b. In and out demand for each CT zone
    c. In and out trip frequency to station level
5. To finalize a list of origin/destination (OD) information by the time interval. For each time interval, there is a set of ODs.

The route choice set is the combination of pick up Citi Bike stations and drop off Citi Bike stations for each OD. The links are not physical road sections in the real world; they are virtual arcs that connect pairs of stations. As congestion occurs in the study network, the congestible capacity effects on a user's choices (e.g. choice of pick up station and drop-off station) should be recognized by the proposed behavior learning model. The shadow prices should reflect stations that become congested with binding capacity effects that result in route diversions.

shows comparison results for the same station using the out-of-sample data set. There's a strong correlation between the model's estimates and its observed values in both plots. The normalized root-mean-square deviation (NRMSD) is computed to compare between observations and model estimates. For station #519, the values of NRMSD are 7.64% and 7.62% for in-sample pick up and drop off, respectively. For out-of-sample, the values of NRMSD are 8.11% and 8.12% for pickup and drop-off, respectively. The lower values of NRMSD (e.g. less than 10%) indicate less residual variance. The average values of NRMSD based on the selected 41 stations in the network are 10.67% (in-sample pick up), 10.62% (in-sample drop off), 15.47% (out-of-sample pick up), and 15.10% (out-of-sample drop off). Values of NRMSD for the full list of stations in the network are shown in Xu *[57]* and posted on the Github site *[52]*.

The estimated capacity functions are applied to obtain congestible capacities. For an observed time interval, non-negative shadow prices are computed if the observed link flow is equal to the estimated capacity (e.g. inbound flow is equal to the number of available docks in the station or outbound flow is equal to the number of available bikes in the station). *R/RStudio 1.1.456* is used to do data processing.

For comparison, route choice probability estimation is run for four scenarios, shown in **Table 3**. For the day of July 18, 2018, the total number of observed trips are 16,940. The basic MNL with no consideration of congestible capacity effects (Model 1) is set as a benchmark, since it is used often in the real world. Model 2 assumes an MNL model where shadow prices are estimated based on an assumed fixed capacity. Model 3 allows for shadow prices estimated from congestible capacity, but with a fixed $\theta$. Model 4 allows $\theta$ to vary among observations.

Table 3. Scenarios evaluated in the case study

| Models | Description |
| --- | --- |
| **Model 1** (Baseline) | MNL with constant link costs ($\theta = 0.1$) |
| **Model 2** | MNL with shadow prices determined whenever the shortest path is not chosen ($\theta = 0.1$) |
| **Model 3** | MNL with congestible capacity effects ($\theta = 0.1$) |
| **Model 4** | MNL with congestible capacity effects ($\theta$ varies among observations) |

For validation, the estimated choices for each scenario are plotted in **Fig. 8**. The moving average of the match score is calculated as a percentage in bold on top of each plot. The baseline model has a match score of 75.32% while the model with congestible capacity effects constraints and constant degree of dispersion has a match score of 77.69%. The model with a variable degree of dispersion has a score that is only 0.8% lower than the model with a constant one, which may be because of the time dimension.

The results clearly suggest two conclusions. The first is that naively assuming a capacity to estimate shadow prices (Model 2) can result in less accurate predictions (vs Model 1). By also incorporating congestible capacities, we see that the model becomes more accurate. Of greater value, however, is that the online model allows us to monitor and quantify the effect that changes in flows have on changes in capacity and their impacts on the consumer surplus of travelers. To demonstrate this online monitoring, we illustrate it for one route from Station 519 (outside Grand Central Station) to Station 362 (37th and Broadway) shown in **Fig. 9**. The model provides a quantitative explanation for the effect of imbalances in bikes and spaces between the two stations on the cost of travel, more than doubling the travel time for the short period or adding up to 50% of the cost shortly prior to 7PM. These costs reflect not just the stations' binding capacities but also those of other stations nearby corresponding to alternative routes and the flows through the system. A modeler using an MNL without the congestible capacity, or for even more accurate predictions using a machine learning model (e.g. *[60]*), but that would not have interpretable results relating these attributes.

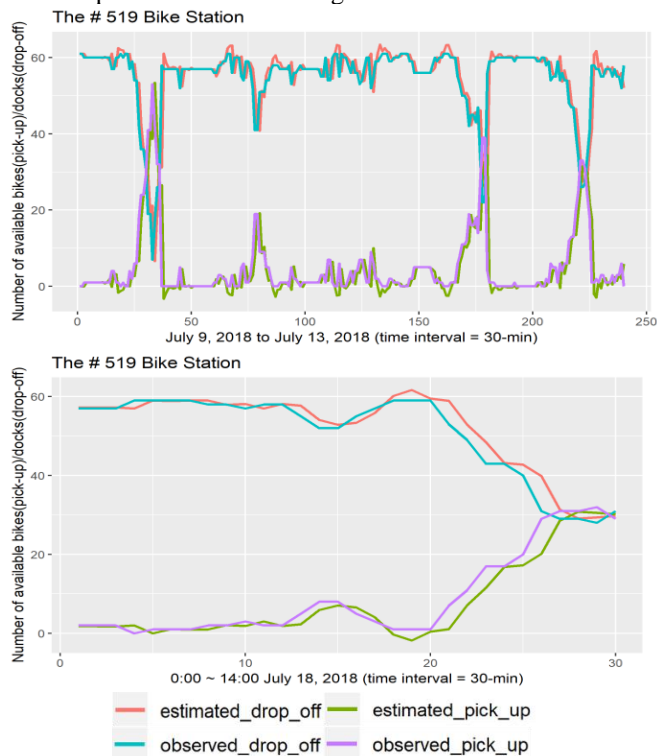

Fig. 7. In-sample and out-of-sample trajectory of the capacities for Citi Bike station #519.

## VI. CONCLUSION AND FUTURE WORK

With the prevalence of MaaS systems, route choice models need to consider characteristics unique to them. MaaS systems tend to involve service systems with fleets of vehicles; as a result, the available service capacity depends on the choices of other travelers in different parts of the system. We model this with a new concept of *congestible capacity*; that is, link capacities rather than link costs are a function of flow. This dependency is also non-separable; the capacity in one link can depend on flows from multiple links.

To model route choice in this setting, a system of offline-estimated equations is used to capture the structural dependency of capacity on inbound and outbound flows, similar to how signalized intersections' capacities depend on arrival patterns of vehicles that can vary from fully random to more platoon-like arrivals. Then, an online-estimated route choice model is used to capture the shadow price corresponding to any binding capacity. This approach of relying on online estimation and observation avoids the endogeneity of the congestible capacities in favor of a practical monitoring and prediction system.

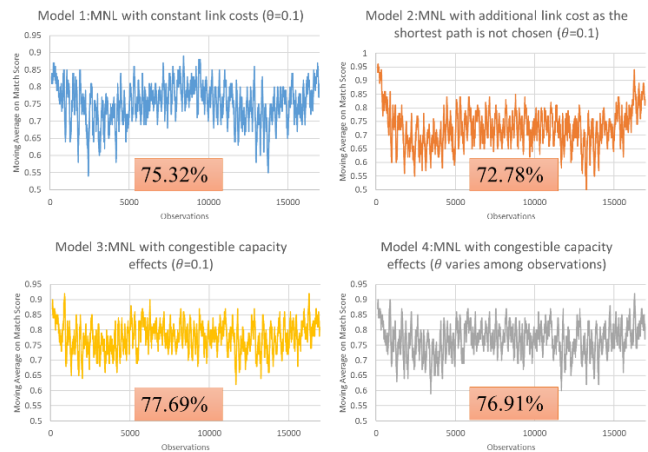

Fig. 8. Comparison results for designed scenarios.

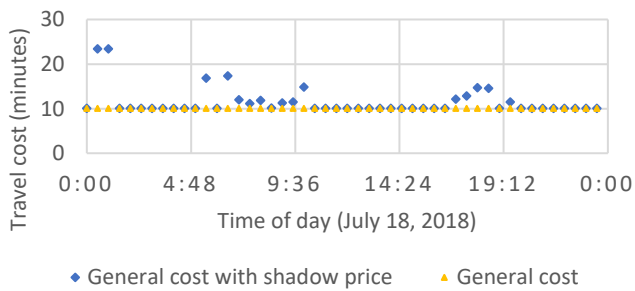

Fig. 9. Travel cost for route from Station 519 to Station 362 by time of day.

The method is first applied to obtain unique congestible capacity shadow prices in a multimodal network. For the numerical experiment, we verify that the methodology has the capability to capture congestion effects on capacities for a multimodal network, where capacities vary and the effects of binding capacities impact the utility of a route. Results show that higher demand lead to higher congestible capacity effects.

The method is validated using real system data from *Citi Bike* in New York City, based on an extracted neighbourhood network in Manhattan midtown, NYC. The results show that the model can fit to the data quite well and performs better than a baseline modeling approach that ignores congestible capacity effects. By relating the route choice to congestible capacities using a random utility model, modelers can monitor and quantify the impacts to traveler consumer surplus in real time.

There are several different avenues for future research. One is to consider a route equilibrium model that is not based on online application, in which a stochastic user equilibration can be obtained between route choice forecasts and assigned flows impacting those choices based on congestible capacity. This may involve the equilibrium model from Brock and Durlauf *[44]* or a mean-field game approach *[53]*. Another avenue is to investigate the use of the online route choice models to support online demand management strategies like customer incentivization programs to help rebalance vehicles (like the Bike Angels program at Citi Bike). A third avenue is to use this approach to estimate real-time route choices in a multimodal setting to dynamically construct path sets in a MaaS network.

This would be useful for developing dynamic MaaS route assignment models. A fourth avenue is to make use of the learning and monitoring aspect for incident management and operations. For example, if a link or node gets disrupted in a time interval, the route choice model can be relied upon to quantify the consumer surplus impacts and anticipate where to allocate resources during the short term. The learning can also be refined to classify travelers by different traveling speeds (walking, bike, driving, etc.) so that route costs can be more accurately estimated for them. The model can be further compared to latent variable models of route choice (e.g. *[58], [59]*) that may capture the latent shadow price variable.

ACKNOWLEDGEMENT

This study was conducted with partial support from NSF grants CMMI-1634973 and CMMI-1652735 and forms one chapter of Susan Jia Xu's PhD dissertation. Helpful comments from Song Gao at UMass Amherst are much appreciated. Any errors and views expressed are solely the authors.

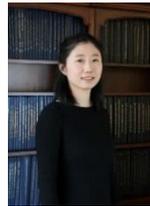

**Susan Jia Xu** received the B.S. degree in civil engineering from Ryerson University, Toronto, Canada, in 2013 and the M.S. degree in transportation system engineering from University of California Irvine, Irvine, CA, USA, in 2015. She got the Ph.D. degree in transportation engineering at New York University, New York, NY, USA.

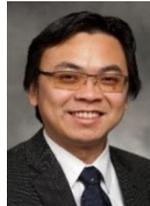

**Joseph Chow** is an Institute Associate Professor in the Department of Civil & Urban Engineering at New York University Tandon School of Engineering, and Deputy Director of the C2SMART University Transportation Center. His research interests lie in emerging mobility in urban public transportation systems, particularly with Mobility-as-a-Service. He obtained his Ph.D. at UC Irvine in 2010, and a BS and MEng at Cornell University in 2000 and 2002.